\RequirePackage{etoolbox}
\csdef{input@path}{{style/}{graphics/}}
\documentclass[ba,linksfromyear,nameyear,dvips,preprint]{imsart}
\usepackage{natbib}
\usepackage[backref=page]{hyperref}
\usepackage{amsthm,amsmath}
\usepackage{graphicx}
\usepackage{algorithm,algorithmic}
\usepackage{multirow}
\usepackage{enumerate}%

\pubyear{2015}
\volume{10}
\issue{1}
\firstpage{1}
\lastpage{30}
\doi{10.1214/14-BA884}% Updated by VTEXPTS2LaTeX.exe, 22.01.2015 14:34

\makeatletter
\newcommand{\bm}[1]{\mbox{\boldmath{$#1$}}}
\makeatother

\begin{document}

\begin{frontmatter}
\title{Bayesian Model Choice in Cumulative Link Ordinal Regression Models}
\runtitle{Bayesian Model Choice in Ordinal Regression Models}

\begin{aug}
\author[a]{\fnms{Trevelyan J.} \snm{McKinley}\corref{}\ead[label=e1]{tjm44@cam.ac.uk}},
\author[a]{\fnms{Michelle} \snm{Morters}\ead[label=e2]{mm675@cam.ac.uk}},
\and
\author[a]{\fnms{James L. N.} \snm{Wood}\ead[label=e3]{jlnw2@cam.ac.uk}}

\runauthor{T. J. McKinley, M. Morters, and J. L. N. Wood}

\address[a]{Disease Dynamics Unit, Department of Veterinary Medicine, University of Cambridge, UK,\\ \printead{e1}}
\end{aug}

%% Abstract %%
\begin{abstract}
The use of the proportional odds (PO) model for ordinal regression is ubiquitous in the literature. If the assumption of parallel lines does not hold for the data, then an alternative is to specify a non-proportional odds (NPO) model, where the regression parameters are allowed to vary depending on the level of the response. However, it is often difficult to fit these models, and challenges regarding model choice and fitting are further compounded if there are a large number of explanatory variables. We make two contributions towards tackling these issues: firstly, we develop a Bayesian method for fitting these models, that ensures the stochastic ordering conditions hold for an arbitrary finite range of the explanatory variables, allowing NPO models to be fitted to any observed data set. Secondly, we use reversible-jump Markov chain Monte Carlo to allow the model to choose between PO and NPO structures for each explanatory variable, and show how variable selection can be incorporated. These methods can be adapted for any monotonic increasing link functions. We illustrate the utility of these approaches on novel data from a longitudinal study of individual-level risk factors affecting body condition score in a dog population in Zenzele, South Africa.
\end{abstract}

%% Keywords %%
\begin{keyword}
\kwd{Bayesian inference}
\kwd{ordinal regression}
\kwd{Markov chain Monte Carlo}
\kwd{reversible-jump}
\kwd{Bayesian model choice}
\end{keyword}

% \begin{keyword}[class=MSC]
% \kwd[Primary ]{}
% \kwd[; secondary ]{}
% \end{keyword}

\end{frontmatter}

%% Mainmatter %%

%s1 ###
\section{Introduction}
The most common regression models for analysing ordinal data fall under the set of cumulative link models, in which the categories of the response variable can be modelled as contiguous intervals on some continuous scale~\citep{mccullagh:1980}. A general monotonic increasing link function is then used to map these intervals from the continuous scale onto the interval $(0,1)$. The choice of link function will generally lead to qualitatively similar model fits, and so can be chosen on the basis of interpretability~\citep{mccullagh:1980} or convenient mathematical properties (e.g.~\citealp{albert_chib:1993}). To this end we concentrate on the logistic link, leading to comparisons of the cumulative odds.

A popular implementation assumes that the relationship between the cumulative log-odds and the explanatory variables does not depend on the response category (the \textit{proportional odds} [PO] model---\citealp{mccullagh:1980}). Under simple constraints, this implementation guarantees that the model exhibits \textit{stochastic ordering} (i.e. it ensures that for $J$ ordered groups, the cumulative probability of belonging to group $j$ is less than or equal to the cumulative probability of belonging to group $j+1$, for $j=1,\dots,J-1$). As a direct result of its simplicity and ease-of-interpretation, the PO model is commonly used in the literature. %However, as with any modelling framework, if the underlying assumptions are invalid, then the model may not fit the data very well and any statistical inference made on the model is compromised~\citep{bender_grouven:1998}.

An alternative is to allow the relationship between the response and explanatory variables to vary by response category (the \textit{non-proportional odds} [NPO] model). Whilst more flexible than the PO model, the use of NPO models in the literature is limited, since the stochastic ordering conditions will only hold for a limited range of values of the explanatory variables \citep{agresti:2010}. This means that many traditional fitting mechanisms can fail to fit~\citep{tutz_scholz:2003}, and in the case where the PO assumption fails to hold, it may be difficult to find estimates for the general NPO model. A useful alternative is to fit a set of $J-1$ separate binary logistic regression models to each cumulative logit separately~\citep{bender_grouven:1998,coleetal:2004}, however, the parameter estimates for the regression parameters do not always correspond to those obtained from the general model~\citep{tutz_scholz:2003}. The \textit{partial proportional odds} (PPO) model~\citep{peterson_harrell:1990} allows for a mixture of PO and NPO variables to be included, though the structure for each variable must be specified in advance. Various approaches have been developed as a means of assessing whether the PO assumption is appropriate for a set of given variables (see e.g.~\citealp{brant:1990,agresti:2010}), but these are difficult to apply to large numbers of variables, particularly if there are interactions between some of them that may impact the relationship.

Another alternative approach---when the response variable is such that to belong to a particular category it is necessary to pass through all previous categories in turn---is to use continuation ratios~\citep{fienberg:1980}. Good reviews of general ordinal regression frameworks can be found in~\cite{ananth_kleinbaum:1997} and~\cite{lalletal:2002}.

The motivating example for this paper is a longitudinal, individual animal-level study of risk factors associated with body condition score in a population of dogs. These data form part of a wider study to examine the impact of immunological and demographic factors on canine rabies vaccination coverage, which covered four locations: Braamfischerville and Zenzele in Gauteng province, South Africa; and Antiga and Kelusa in Bali province, Indonesia. Full details of the wider study, and a comprehensive analysis of all the data collected from each of the sites is provided in~\cite{mortersetal:2014}.

To illustrate the requirement and performance of the methodology, we focus attention on one particular data set from Zenzele. Full details of these data are given in Section~\ref{sec:app}. The response variable is body condition score (BCS)---defined on a scale from 1--9 where a score of 1 is highly underweight, 5 is healthy, and 9 is highly overweight. As such it seems sensible to consider using cumulative link models.

There are various challenges with modelling this system, and we expand on each point in the subsequent discussion:
\begin{enumerate}
    \item We wish to perform variable selection, in order to assess the relative impact and importance of a series of potential risk factors on BCS.
    \item The data are longitudinal, and so it is necessary to account for clustering due to repeated measurements on individual animals.
    \item We wish to assess the weight-of-evidence for PO versus NPO structures for each of the variables. This information is useful in helping to build up a picture (along with other indirect sources of evidence---see~\citealp{mortersetal:2014}) of the environmental processes driving canine demography in these regions (see Section~\ref{sec:app} for a more detailed discussion of this point).\label{item3}
    \item In order to tackle point~\ref{item3}, it is necessary to overcome some of the challenges regarding the fitting of NPO models when the stochastic ordering conditions may not hold.
\end{enumerate}
In a classical statistical framework, model choice is usually performed using some form of model comparison criteria, such as Akaike's Information Criterion (AIC;~\citealp{akaike:1974}), or likelihood ratio testing. These procedures use information from a single point estimate of the parameters, $\hat{\bm{\theta}}$, and neglect the uncertainty in $\bm{\theta}$. In addition, inference is made conditionally on the selected model, and does not incorporate uncertainty in the choice of model. This can be important in cases where explanatory variables show consistent evidence of an effect across a range of models, but is not selected in the `final' model (see e.g.~\citealp{viallefontetal:2001}). Here we propose to use a Bayesian framework, and implement model selection using posterior probabilities of association (see e.g.~\citealp{kass_raftery:1995,viallefontetal:2001,ohara_sillanpaa:2009}). This has the advantage that it allows us to assess the weight-of-evidence in favour of a given model, and also allows us to assess the evidence for a particular variable being associated with the response after averaging across all possible models. This is particularly important to this study, since we also wish to assess the conditional evidence of a PO or NPO structure for the relationship between an explanatory variable and the response, given that an association exists.

A common method to account for clustering due to repeated measurements is to use mixed effects models (see e.g.~\citealp{diggleetal:2002}), in which the error term is split into different components in order to model the variance-covariance structure at different hierarchies (\citealp{hedeker_gibbons:1994,gibbons_hedeker:1997,hartzeletal:2001,hedeker:2003,liu_hedeker:2006}). These techniques are well characterised in the literature, though there is some debate about how to perform model selection in the presence of random effects in a classical setting (see e.g.~\citealp{vaida_blanchard:2005,liangetal:2008}). In the Bayesian setting, all parameters are considered random variables, and it is straightforward to incorporate \textit{a priori} clustering into the prior. The model choice problem then remains the same, as the parameters are simply integrated over when estimating the posterior probabilities of association.

The literature surrounding the development of ordinal regression frameworks is large and varied, applied in a wide range of fields. Focussing on Bayesian models; probit link functions are frequently used (e.g.~\citealp{albert_chib:1993,chu_ghahramani:2005,yietal:2007}) and there are various recent developments in modelling the link function using mixture distributions (e.g.~\citealp{lang:1999,leonnoveloetal:2010}). Different fitting mechanisms have also been developed, including Markov chain Monte Carlo (MCMC---\citealp{lang:1999,ishwaran_gatsonis:2000,holmes_held:2006,yietal:2007,webb_forster:2008,leonnoveloetal:2010}), Laplace approximations (e.g. \citealp{chu_ghahramani:2005};~\citealp{paquetetal:2005}) and expectation-propagation algorithms (\citealp{chu_ghahramani:2005}).~\cite{holmes_held:2006} develop efficient MCMC samplers for logistic multinomial regression models, and~\cite{obrien_dunson:2004} develop a multivariate logistic regression framework that provides a marginal logistic structure for each of the outcomes. Some work has also been done on model selection using probit models (e.g.~\citealp{albert_chib:1997,chu_ghahramani:2005,webb_forster:2008}), and~\cite{mwalilietal:2005} extend a PO model to account for interobserver measurement error. This list is by no means exhaustive, but as far as we are aware no method has been developed that accounts for all four of the challenges we highlighted earlier within the same framework. This manuscript is an attempt to provide an alternative approach for fitting logistic regression models, which allows both PO and NPO structures to be used (and provides a model-driven means of assessing which structure is most relevant for each variable in the presence of other variables), and which can be extended to deal with repeated or clustered measurements, as well as variable selection.

The first challenge we address is to provide a framework in which the stochastic ordering conditions can be made to hold for any given data set. This provides a means to explore the fitting of NPO models to any data set, and facilitates the development of a more general approach in which the relationship between the response and explanatory variables (e.g. PO or NPO) can be allowed to vary according to the data. The latter is our second contribution, and is useful because often we do not know which explanatory variables are best modelled using PO or NPO structures in advance, particularly when there are a large number of variables. To this end,~\cite{tutz_scholz:2003} propose a method that switches between the PO, PPO and NPO models, fitting the model via a penalised likelihood approach. However, in this case we also would like to produce an estimate of the support under the data for these competing structures for each of the variables, which can provide some indirect evidence regarding the mechanisms at play in the underlying system. Although the PO model could be viewed as a special case of the NPO model, the na\"{i}ve use of a straight NPO model could result in overfitting.
%In addition, if the stochastic ordering conditions do not hold, it is not possible to fit the NPO model. In Section~\ref{sec:criteria} we provide a means to ensure the stochastic ordering conditions hold for a given data set.

The challenge with comparing PO and NPO structures is that the dimensionality of the system is different in each case (a single regression parameter for the PO model corresponds to $J-1$ parameters for the NPO model). To deal with this issue we use reversible-jump Markov chain Monte Carlo (RJ-MCMC---\citealp{green:1995}) to fit the model, and Bayesian model averaging (BMA---e.g.~\citealp{kass_raftery:1995}) to produce posterior probabilities of association (PPAs) for the support under the prior and the data for the choice of PO or NPO structure, averaged across the set of possible models. Finally we extend these ideas to incorporate variable selection (see e.g.~\citealp{dellaportasetal:2002,ohara_sillanpaa:2009}). Note that an implementation of model choice based on Bayes Factors for ordinal regression models was developed in~\cite{albert_chib:1997}, though each competing model needs to be fitted separately in order to be compared~\citep{chib:1995}. Here we integrate across the competing models in one framework, which is likely to be much more efficient when searching across a large model space. An alternative Bayesian RJ-MCMC approach to model choice for ordinal probit models is developed in~\cite{webb_forster:2008}.

In Section~\ref{sec:bayes} we discuss the Bayesian paradigm and (RJ-)MCMC. In Section~\ref{sec:cumlink} we introduce the general cumulative link model, and more specifically the PO, NPO and PPO models. In Section~\ref{sec:NPO} we discuss how the Bayesian framework can be used to ensure stochastic ordering for specified variable ranges, and we justify this approach in practice in Section~\ref{sec:criteria}. The specific RJ-MCMC sampler for this ordinal model choice problem is described in Section~\ref{sec:RJ}. We then apply these methods to both simulated data, as well as data from a longitudinal study of individual-level risk factors affecting body condition score in a dog population in Zenzele, South Africa (Section~\ref{sec:app}). We conclude with a discussion (Section~\ref{sec:dis}).

%s2 ###
\section{Bayesian inference and Markov chain Monte Carlo}
\label{sec:bayes}
All of the models described in this paper will be formulated in a Bayesian framework, and fitted using Markov chain Monte Carlo (MCMC). We assume that readers are familiar with the Bayesian framework, but otherwise they are referred to various excellent texts available, such as~\cite{gilksetal:1996,gelmanetal:2004} and~\cite{gamerman_lopes:2006}. The model fitting algorithms described in this manuscript are specifically variations of the Metropolis-Hastings (M-H) algorithm~\citep{metropolisetal:1953,hastings:1970}.

Reversible-jump MCMC~\citep{green:1995} is an extension to the classic M-H routine that allows the Markov chain to jump between models with different dimensionality. Again, we do not discuss the full details of RJ-MCMC here, but for good introductions to the method the reader is referred to papers by~\cite{waagepetersen_sorensen:2001} and~\cite{hastie_green:2012}.

%To implement model choice in an RJ-MCMC framework, we simply introduce the model indicator $v$ as an additional parameter in the system. The RJ algorithm takes a similar form to the traditional M-H algorithm, except that it also allows for moves between competing models.
%In order to implement RJ-MCMC, we allow the Markov chain to propose jumps between competing models at each iteration.

%s2.1 ###
\subsection{Bayesian model choice using reversible-jump MCMC}
\label{sec:bayeschoice}
Assume that we have $V$ competing models to choose between. We can formulate the Bayesian model choice problem as one of estimating the posterior probability that a model ($M_v$) is true, given the choice of competing models ($M_1,\dots,M_V$). Formally, this quantity is defined as
%e1 ###
\begin{equation}
\label{eq:PPA}
    P\left(M_v \mid \bm{D} \right) = \frac{f\left(\bm{D} \mid M_v\right)P\left(M_v\right)}{\sum_{u=1}^V f\left(\bm{D} \mid M_u\right)P\left(M_u\right)},
\end{equation}
with $P\left(M_v\right)$ the \textit{prior} probability of association for model $M_v$, and
%e2 ###
\begin{equation}
\label{eq:intlike}
    f\left(\bm{D} \mid M_v\right) = \int_{\bm{\Omega}_v} f\left(\bm{D} \mid \bm{\omega}_v, M_v\right)f\left(\bm{\omega}_v \mid M_v\right) d\bm{\omega}_v,
\end{equation}
the \textit{integrated likelihood}; where $\bm{D}$ is the data, and $\bm{\Omega}_v$ is the (multidimensional) parameter space for the unknown parameters $\bm{\omega}_v$ corresponding to model $M_v$. The quantity (\ref{eq:PPA}) is sometimes referred to as the \textit{posterior probability of association} (PPA). These ideas for Bayesian model choice go back originally to~\cite{jeffreys:1935,jeffreys:1961}, and for a detailed introduction see~\cite{kass_raftery:1995} and~\cite{ohara_sillanpaa:2009}.

To implement model choice in an RJ-MCMC framework, let $v = i$ be the model indicator at a given iteration (i.e. the chain is in model $M_i$), then let $p\left( M_i \to M_j \right)$ denote the probability that a jump from $M_i$ to $M_j$ is proposed. In order to jump between models of differing dimensionality, the parameters $\bm{\omega}_i$ are mapped to a set of parameters $\bm{\omega}_j$ via the inclusion of a set of dummy parameters, $\bm{u}_i$ and $\bm{u}_j$, that are chosen to ensure that $\mbox{dim}\left(\bm{\omega}_i,\bm{u}_i\right)=\mbox{dim}\left(\bm{\omega}_j,\bm{u}_j\right)$. Once these dummy parameters are chosen, $\left(\bm{\omega}_i,\bm{u}_i\right)$ is mapped to $\left(\bm{\omega}_j,\bm{u}_j\right)$ through a deterministic bijective function $g_{ij}$, such that    $g_{ij}\left(\bm{\omega}_i,\bm{u}_i\right) = \left(\bm{\omega}_j,\bm{u}_j\right)$, and the reverse move is $g_{ji}\left(\bm{\omega}_j,\bm{u}_j\right) = \left(\bm{\omega}_i,\bm{u}_i\right)$. The acceptance probability of the move is then given by:
%e3 ###
\begin{eqnarray}
\label{eq:RJacc}
    &&\alpha = \min \left[ 1,\frac{f\left(\bm{D} \mid \bm{\omega}_j,M_j\right)}{f\left(\bm{D} \mid \bm{\omega}_i, M_i\right)}
     \times \frac{f\left(\bm{\omega}_j \mid M_j\right)}{f\left(\bm{\omega}_i \mid M_j\right)}
     \times \frac{P\left(M_j\right)}{P\left(M_i\right)}\right. \nonumber \\
     && \left.\qquad\qquad\qquad\qquad
    \times \frac{p(M_j \to M_i)q_u\left(\bm{u}_j\right)}{p(M_i \to M_j)q_u\left(\bm{u}_i\right)}
    \times \left|\frac{\partial \left( \bm{\omega}_j,\bm{u}_j\right)}{\partial \left( \bm{\omega}_i,\bm{u}_i\right)} \right|
        \right],
\end{eqnarray}
where $q_u\left(\bm{u}_i \right)$ is the proposal density for the dummy parameters $\bm{u}_i$, and likewise for $q_u\left(\bm{u}_j\right)$. The final quantity in (\ref{eq:RJacc}) is the absolute value of the determinant of the Jacobian matrix.

One advantage of using RJ methodology is that for a well-mixing model, the PPA defined in (\ref{eq:PPA}) for a model $M_v$ can be simply estimated as the proportion of time that the chain spends in model $v$. In Section~\ref{sec:RJ} we show how this routine can be implemented for variable selection, as well as choosing between PO and NPO structures for individual variables. For other applications of RJ-MCMC in Bayesian model choice see e.g.~\cite{richardson_green:1997} and~\cite{dellaportasetal:2002}.

\section{Cumulative link models}
\label{sec:cumlink}
Let $Y=\left(Y_{1},\dots,Y_{J}\right)$ be a set of counts of individuals in $j=1,\dots,J$ \textit{ordered} categories, which can be modelled as
%e4 ###
\begin{equation}
    Y \sim \mbox{Mult}\left(n,\bm{p}\right),
\end{equation}
where $n$ is the number of individuals, and $\bm{p}=\left(p_{1},\dots,p_{J}\right)$ correspond to the probabilities of each individual being in any given category $j$ (such that $\sum_{j=1}^J p_j=1$). If we have a set of $K$ explanatory variables, $\bm{X}_i=\left(X_{i1},\dots,X_{iK}\right)$, associated with subset $i$ of the $n$ individuals (where $i=1,\dots,I$, such that $n=n_1 + \dots + n_I$), then
%e5 ###
\begin{equation}
\label{eq:multinom}
    Y_i \sim \mbox{Mult}\left(n_i,\bm{p}_i\right),
\end{equation}
where $\bm{p_i}=\left(p_{i1},\dots,p_{iJ}\right)$ and $\sum_{j=1}^J p_{ij}=1$. For a fully individual-based model then $I=n$ and $n_i=1$ for all $i$. Letting $C_i$ correspond to the category that an individual $i$ belongs to (such that $C_i$ takes values $1,\dots,J$), then following~\cite{mccullagh:1980}, we can model the cumulative probabilities, $P\left(C_i\leq j\right)=\gamma_{ij}$ through a monotonic increasing \textit{link} function $h(\cdot)$, mapping the interval $(0,1) \to (-\infty,\infty)$, as
%e6 ###
\begin{equation}
\label{eq:link}
    h\left(\gamma_{ij}\right)=\theta_j-\mu_i,
\end{equation}
where $\mu_i=\beta_0 + \beta_1 X_{i1} + \dots + \beta_K X_{iK}$ is a linear regression term, and $\bm{\beta}=\left(\beta_0,\dots,\beta_K\right)$ is a vector of $K+1$ regression parameters. In this framework the $\theta_j$ parameters correspond to a set of latent continuous `cut-points', such that $-\infty < \theta_1 < \dots < \theta_{J-1} < \infty$. For identifiability we set $\beta_0=0$. The probabilities of category membership are then given by
%e7 ###
\begin{equation}
\label{eq:ps}
    p_{ij}=\left\{ \begin{array}{cl}
            \gamma_{ij} & \mbox{for $j=1$},\\
            \gamma_{ij}-\gamma_{i(j-1)} & \mbox{for $j=2,\dots,J-1$},\\
            1-\gamma_{i(J-1)} & \mbox{for $j=J$}.
            \end{array}
            \right.
\end{equation}
In (\ref{eq:link}), the effect of the explanatory variables is independent of the grouping, and so regardless of the choice of link function the models display \textit{strict stochastic ordering}~\citep{mccullagh:1980}. This means that subject to the constraint $-\infty < \theta_1 < \dots < \theta_{J-1} < \infty$, the cumulative probabilities will be such that $0<\gamma_{i1} < \gamma_{i2} < \dots < \gamma_{i(J-1)} < 1$.

%As noted by~\cite{mccullagh:1980}, in practice this property means that all linear models with the structure (\ref{eq:link}) are likely to be qualitatively similar, and so suggests that the choice of link function should be made based on ease-of-interpretation, though in many cases the choice of link function is also made based on convenient mathematical properties, such as ease-of-sampling (\citealp{albert_chib:1993}).

A more general model would allow the effect of the covariates to vary between the groups, such that
%e8 ###
\begin{equation}
    h\left(\gamma_{ij} \right) = \theta_j - \mu_{ij}.
\end{equation}
In this case the models are only stochastically ordered for certain ranges of explanatory variables (see e.g.~\citealp{agresti:2010,congdon:2005,tutz_scholz:2003}). We discuss both forms of these models in detail for the case of the logistic-link function, but also extend the discussion to more general cases.

%s3.1 ###
\subsection{Proportional odds model}
\label{sec:PO}
A common form for the link function is the \textit{logistic} link:
%e9 ###
\begin{equation}
\label{eq:propodds}
    h\left(\gamma_{ij}\right)=\log\left(\frac{\gamma_{ij}}{1-\gamma_{ij}}\right)=\theta_j-\mu_i.
\end{equation}
This is known as the \textit{proportional odds} (PO) model~\citep{mccullagh:1980}, so-called because the cumulative log-odds ratio for two sets of explanatory variables, $\bm{X}_1$ and $\bm{X}_2$ is given by
%e10 ###
\begin{eqnarray}
    \mbox{logit}\left(\gamma_{1j}\right)-\mbox{logit}\left(\gamma_{2j}\right) &=& \theta_j - \bm{\beta}^T \bm{X}_1 - \theta_j + \bm{\beta}^T \bm{X}_2\nonumber \\
    &=& \bm{\beta}^T\left(\bm{X}_2-\bm{X}_1\right).
\end{eqnarray}
Hence the cumulative log-odds ratio is proportional to the distance between $\bm{X}_1$ and $\bm{X}_2$ (see also~\citealp{agresti:2010}).

In the case of the PO model (\ref{eq:propodds}), the $\bm{\theta}$ and $\bm{\beta}$ parameters are \textit{a priori} independent, and so the joint prior distribution can be written as $f\left(\bm{\beta},\bm{\theta}\right) = f\left(\bm{\theta}\right)f\left(\bm{\beta}\right)$, where we let
%e11 ###
\begin{equation}
    f\left(\bm{\beta}\right) = \prod_{k=1}^K f\left(\beta_k\right)~\mbox{and}~f\left(\bm{\theta}\right) = f\left( \theta_1\right)\prod_{j=2}^{J-1} f\left( \theta_j \mid \theta_{j-1}\right),
\end{equation}
where $f\left(\theta_1\right)$ is defined in $(-\infty,\infty)$, and $f\left(\theta_j \mid \theta_{j-1}\right)$ is defined in the range $\left(\theta_{j-1},\infty\right)$ for $j=2,\dots,J-1$ (see also~\citealp{albert_chib:1993,johnson_albert:1999,congdon:2005}). This ensures stochastic ordering for any values of $\bm{\beta}$. We let $\theta_1 \sim N\left(0,\sigma_{\theta}^2\right)$, $\beta_k \sim N\left(0,\sigma_{\beta}^2\right)$ (for $k=1,\dots,K$) and
%e12 ###
\begin{equation}
    \theta_j \mid \theta_{j-1} \sim N\left(0,\sigma_{\theta}^2\right)I\left(T_{j-1},\infty\right)~\mbox{for $j=2,\dots,J-1$},
\end{equation}
where $I\left(T_{j-1},\infty\right)$ signifies that the distribution is truncated in the region $\left(T_{j-1},\infty\right)$ (i.e. it is a \textit{lower-truncated} normal distribution) with $T_{j-1}=\theta_{j-1}$.
%This has probability density function given by
%   f\left(\theta_j \mid \theta_{j-1} \right) = \frac{\frac{1}{\sigma_{\theta}\sqrt{2\pi}}\exp\left(-\frac{\theta_j^2}{2\sigma_{\theta}^2}\right)}{1-\Phi\left(\theta_{j-1}\right)},
%where, $\Phi(\cdot)$ is the cumulative distribution function for a normal distribution with mean 0 and variance $\sigma_{\theta}^2$.
Other alternative choices for the prior distributions include doubly-truncated normals~\citep{congdon:2005}, an ordered uniform distribution~\citep{ishwaran:2000}, or a re-parameterisation which maps the constrained variables $\bm{\theta}$ to a set of unconstrained variables, $\bm{\alpha}$, which can be given, for example, a multivariate normal prior~\citep{fahrmeier_tutz:1994,albert_chib:1997}. We choose normal random walk proposal distributions for each $\beta_k$, such that
%e13 ###
\begin{equation}
    \beta_k' \mid \beta_k^{(i)} \sim N\left(\beta_k^{(i)}, \sigma_{P\beta}^2\right),
\end{equation}
where $\sigma_{P\beta}^2$ is the proposal variance. For the cut-point parameters, $\theta_j$, we choose truncated uniform random-walk proposals, such that
%e14 ###
\begin{equation}
    \theta_j' \mid \bm{\theta}^{(i)} = \left\{ \begin{array}{ll}
            U\left(\theta_j^{(i)}-\tau_{\theta},\min\left[\theta_j^{(i)}+\tau_{\theta},\theta_{j+1}^{(i)}\right]\right) & \mbox{if $j=1$},\\[6pt]
            U\left(\max\left[\theta_j^{(i)}-\tau_{\theta},\theta_{j-1}^{(i)}\right],\min\left[\theta_j^{(i)}+\tau_{\theta},\theta_{j+1}^{(i)}\right]\right) & \mbox{if $j=2,\dots,J-2$},\\[6pt]
            U\left(\max\left[\theta_j^{(i)}-\tau_{\theta},\theta_{j-1}^{(i)}\right],\theta_j^{(i)}+\tau_{\theta}\right) & \mbox{if $j=J-1$},\\
            \end{array}
            \right.
\end{equation}
where $\tau_{\theta}>0$ controls the size of the maximum unconstrained move away from the current value at each iteration.

%s3.2 ###
\subsection{Non-proportional odds model}
\label{sec:NPO}
The non-proportional odds (NPO) model is specified as
%e15 ###
\begin{equation}
\label{eq:nonpropodds}
    \log\left(\frac{\gamma_{ij}}{1-\gamma_{ij}}\right)=\theta_j-\mu_{ij}.
\end{equation}
In this case the regression parameters are allowed to vary with category level, such that $\mu_{ij}=\bm{\beta}_j^T\bm{X}_i$ (see e.g.~\citealp{agresti:2010,bender_grouven:1998,tutz_scholz:2003,congdon:2005}). The key challenge is that in order for stochastic ordering to hold, it is necessary that
%e16 ###
\begin{equation}
\label{eq:stochorder}
    -\infty < \theta_1-\bm{\beta}_1 \bm{X} < \theta_2-\bm{\beta}_2 \bm{X} < \dots < \theta_{J-1}-\bm{\beta}_{J-1} \bm{X} < \infty
\end{equation}
for all $\bm{X}$. For identifiability we set each of the intercept parameters $\beta_{0j}=0$. If we have $K$ explanatory variables, then after expanding out the regression in the stochastic ordering constraints (\ref{eq:stochorder}), for any $j=1,\dots,J-2$, we have that
%e17 ###
\begin{equation}
\label{eq:nonPOcond}
    \theta_j - \theta_{j+1} < \sum_{k=1}^K \left(\beta_{kj} - \beta_{k(j+1)}\right)X_k,
\end{equation}
which must hold for any value of $X_k$.

In the first instance, assume that we have a lower and upper bound for the possible values of $X_k$, such as would be the case if $X_k$ were categorical. Denote the minimum and maximum values of $X_k$ as $X_k^m$ and $X_k^M$ respectively. The condition (\ref{eq:nonPOcond}) then becomes
%e18 ###
\begin{equation}
\label{eq:nonPOcond2}
    \theta_j - \theta_{j+1} < \sum_{k=1}^K \min\left(\left[\beta_{kj} - \beta_{k(j+1)}\right]X_k^m,\left[\beta_{kj} - \beta_{k(j+1)}\right]X_k^M \right).
\end{equation}
For brevity, let
%e19 ###
\begin{equation}
\label{eq:Ckj}
    C_{kj} = \min\left[X_k^m\left(\beta_{kj} - \beta_{k(j+1)}\right),X_k^M\left(\beta_{kj} - \beta_{k(j+1)}\right)\right]~\mbox{and}~C_j = \sum_{k=1}^K C_{kj}.
\end{equation}

In a similar manner to the PO model, we can therefore specify the joint prior distribution of $\bm{\beta}$ and $\bm{\theta}$; however this time we do not assume both sets of parameters are independent, hence
%e20 ###
\begin{eqnarray}
\label{eq:nonpropprior}
    f\left(\bm{\beta},\bm{\theta}\right) &=& f\left(\bm{\theta} \mid \bm{\beta} \right)f\left(\bm{\beta}\right) \nonumber\\
    &=& f\left(\theta_1 \right) \left[\prod_{j=2}^{J-1} f\left(\theta_j \mid \theta_{j-1},\bm{\beta}_{j},\bm{\beta}_{j-1} \right)\right]\prod_{k=1}^K f\left(\beta_k\right),
\end{eqnarray}
where $f\left(\beta_k\right)$ ($k=1,\dots,K$) and $f\left(\theta_1\right)$ are defined as before, and\\ $f\left(\theta_j \mid \theta_{j-1},\bm{\beta}_{j},\bm{\beta}_{j-1} \right)$ is the probability density function for a truncated normal distribution, $N\left(0,\sigma_{\theta}^2\right)I\left(T_{j-1},\infty\right)$ with $T_{j-1}=\theta_{j-1}-C_{j-1}$. We choose to update each $\beta_{kj}$ parameter in turn, conditional on all other parameters remaining fixed. It is tricky to define a simple mechanism for truncated sampling of the regression parameters, due to the fact that the conditions (\ref{eq:stochorder}) change according to whether we propose $\beta_{kj}'<\beta_{k(j+1)}^{(i)}$ or $\beta_{kj}'>\beta_{k(j+1)}^{(i)}$. Instead we opt here for a simple random-walk proposal, such that
%e21 ###
\begin{equation}
    \beta_{kj}' \mid \beta_{kj}^{(i)} = U\left(\beta_{kj}^{(i)}-\tau_{\beta},\beta_{kj}^{(i)}+\tau_{\beta}\right),
\end{equation}
where $\tau_{\beta}>0$ controls the size of the maximum move away from the current value at each iteration. For the cut-point parameters, $\theta_j$, we choose truncated uniform random-walk proposals, such that
%e22 ###
\begin{equation}
    \theta_j' \mid \bm{\theta}^{(i)} = \left\{ \begin{array}{ll}
            U\left(\theta_j^{(i)}-\tau_{\theta},\min\left[\theta_j^{(i)}+\tau_{\theta},\theta_{j+1}^{(i)}+C_j^{(i)}\right]\right) & \mbox{if $j=1$},\\[6pt]
            U\left(\max\left[\theta_j^{(i)}-\tau_{\theta},\theta_{j-1}^{(i)}-C_{j-1}^{(i)}\right],\right.&\\[6pt]
            \qquad \left.\min\left[\theta_j^{(i)}+\tau_{\theta},\theta_{j+1}^{(i)}+C_j^{(i)}\right]\right) & \mbox{if $j=2,\dots,J-2$},\\[6pt]
            U\left(\max\left[\theta_j^{(i)}-\tau_{\theta},\theta_{j-1}^{(i)}-C_{j-1}^{(i)}\right],\theta_j^{(i)}+\tau_{\theta}\right) & \mbox{if $j=J-1$},\\
            \end{array}
            \right.
\end{equation}
where $\tau_{\theta}>0$ controls the size of the maximum unconstrained move.

%s3.3 ###
\subsection{Partial proportional odds}
The \textit{partial proportional odds} (PPO) model, proposed by~\cite{peterson_harrell:1990}, allows some variables to have a proportional odds structure and some to not. It takes the form
%e23 ###
\begin{equation}\label{eq:PPOeq}
    \log\left(\frac{\gamma_{ij}}{1-\gamma_{ij}}\right) = \theta_j-\bm{\beta}^T\bm{X}_i - \bm{\eta}_j^T\bm{U}_i,
\end{equation}
where $i=1,\dots,n$ and $j=1,\dots,J-1$. The regression parameters $\bm{\beta}$ correspond to the set of explanatory variables, $\bm{X}_i$, that have a proportional odds structure, and the regression parameters $\bm{\eta}_j$ correspond to the set of explanatory variables, $\bm{U}_i$, that have a non-proportional odds structure.
%~\cite{peterson_harrell:1990} also proposed a \textit{constrained} version of this model, where the parameters $\bm{\eta}_j$ are subjected to constraints, usually informed by the observed data.
The approaches described in Sections~\ref{sec:PO} and~\ref{sec:NPO} can be combined in order to fit a PPO model, where the PO and NPO variables ($\bm{X}$ and $\bm{U}$) are known in advance. In subsequent sections we formulate an approach whereby the optimal choice of PO or NPO structure for each explanatory variable can instead be directly estimated by the model.

%s3.4 ###
\subsection{Justification of approach}
\label{sec:criteria}
The approach described in the previous section assumes that each of the $X_k$ variables is bounded in some finite region, which is true for any set of categorical explanatory variables, since for a categorical variable $Z_i$ with $L$ levels ($0,\dots,L-1$), it is straightforward to represent $Z_i$ as a set of $L-1$ dummy variables, $X_{i1}, \dots, X_{i(L-1)}$, such that
%e24 ###
\begin{equation}
    X_l = \left\{ \begin{array}{ll}
            1 & \mbox{if $Z_i=l$},\\
            0 & \mbox{otherwise}.
        \end{array}
        \right.
\end{equation}
In the case of continuous or discrete explanatory variables that are bounded in a finite range, then the approach described in Section~\ref{sec:NPO} will also ensure stochastic ordering holds. However, if $X_k$ is defined over an infinite range, then these conditions will break for some values of $X_k$ if the $\beta_{kj}$ parameters are also unbounded.

Here we argue for a pragmatic solution to this problem, by considering that it is possible to define an upper and lower bound for $X_k$ based on the observed data, and then use (\ref{eq:nonPOcond2}) to set boundary conditions for the conditional priors in (\ref{eq:nonpropprior}). Although this does not mean that the stochastic ordering will hold for all possible theoretical values of $X_k$, it does ensure that the stochastic ordering will hold for the range of values found in the observed data. We make two main arguments to justify this approach:
\begin{enumerate}
    \item Although theoretically the values for $X_k$ might be infinite, for any practical applications of the model there will almost certainly be a finite range of possible values. If the observed data are a fair representation of the underlying population, then provided the model is a good fit, any population-level inference made from the model is likely to be fairly robust (i.e. the posterior distribution is the distribution of the parameters \textit{given} the observed data, so this is explicitly represented within the Bayesian paradigm).
    \item The assumptions underlying \textit{any} statistical model can only be assessed across the range of values used to fit the data, there is no guarantee that the assumptions will hold beyond this range, even if it is possible to extrapolate without breaking any conditions of the model.
\end{enumerate}

%s4 ###
\section{Reversible-jump algorithm for variable selection in cumulative odds ordinal regression models}
\label{sec:RJ}
Consider that we have $K$ parameters describing the explanatory variables. In the first instance assume that each parameter measures the effect of a single variable (i.e. there are no categorical variables with $>2$ levels, or any interaction effects). We will extend discussion to these more complex variables in due course. We can model the relationship between the response variable $Y$ and each variable $X_k$ in one of three ways: either with a PO structure, an NPO structure or no relationship at all; in this example giving $3^K$ possible models. Here we will assume that we have no prior information to distinguish between which of these models is most likely, and so assume equal prior probabilities of association for each competing model. To model these structures we introduce an indicator variable $S_k$, for $k=1,\dots,K$, where
%e25 ###
\begin{equation}
    S_k = \left\{ \begin{array}{ll}
        0 & \mbox{if $X_k$ has a PO structure},\\
        1 & \mbox{if $X_k$ has an NPO structure},\\
        2 & \mbox{if $X_k$ is excluded}.
    \end{array}
    \right.
\end{equation}
To ease programming, it is helpful to treat each of these three possibilities as special cases of the NPO-structure, such that if a variable has a PO-structure then this is equivalent to setting $\beta_{kj}=\beta_{k}$ for $j=1,\dots,J-1$, with independent point-mass priors on $\beta_{k2},\dots,\beta_{k(J-1)}$ such that $f\left(\beta_{kj}=0\right)=1$. If a variable is excluded then this is equivalent to setting $\beta_{kj}=0$ with a point-mass prior $f\left(\beta_{kj}=0\right)=1$. This enables us to use the conditions in (\ref{eq:stochorder}) to ensure general stochastic ordering.

%s4.1 ###
\subsection{Adding or removing variables}
\label{sec:addrem}
Our stochastic search routine updates each variable $X_k$ in a random order, by proposing to add the variable (if currently excluded) or to remove the variable (if currently included) with a probability $p_{\mbox{\scriptsize jump}}$ (hence we do nothing with a probability $1-p_{\mbox{\scriptsize jump}}$).

To add a variable into the model, we sample whether to use a PO or NPO structure with probability $p_{\mbox{\scriptsize PO}}$ and $p_{\mbox{\scriptsize NPO}}=1-p_{\mbox{\scriptsize PO}}$ respectively. To add a variable to the model with a PO structure, we define a bijective function
%e26 ###
\begin{equation}
\label{eq:gzerotoPO}
    g_{0 \to PO} \left(u_1\right) = u_1=\beta_k,
\end{equation}
where $u_1$ is sampled from some distribution with p.d.f. $q_u(\cdot)$. To add a variable with an NPO structure, we define a bijective function
%e27 ###
\begin{equation}
\label{eq:gzerotoNPO}
    g_{0 \to NPO} \left(u_1,\dots,u_{J-1}\right) = \left(u_1,\dots,u_{J-1}\right)=\left(\beta_{k1},\dots,\beta_{k(J-1)}\right),
\end{equation}
where $u_1,\dots,u_{J-1}$ are independent and identically distributed (i.i.d.) samples from a distribution following $q_u(\cdot)$. For a 0 $\to$ PO move, the acceptance probability
is\vadjust{\goodbreak}
%e28 ###
\begin{equation}
    \alpha = \min \left[ 1,\frac{f\left(\bm{Y} \mid \bm{\beta}',\bm{\theta}^{(i)}\right)}{f\left(\bm{Y} \mid \bm{\beta}^{(i)},\bm{\theta}^{(i)}\right)}
     \times \frac{f\left(\beta_k'\right)}{1} \times \frac{1}{q_u\left(u_1\right)}\times \frac{1}{p_{\mbox{\scriptsize PO}}}\right].
\end{equation}
The probability of adding or dropping a variable, $p_{\mbox{\scriptsize jump}}$, is the same for the forwards and reverse moves, and so cancel in the acceptance ratio. The determinant of the Jacobian matrix is 1. For a 0 $\to$ NPO move, the acceptance probability is
%e29 ###
\begin{eqnarray}
    &&\alpha = \min \Biggl[ 1,\frac{f\left(\bm{Y} \mid \bm{\beta}',\bm{\theta}^{(i)}\right)}{f\left(\bm{Y} \mid \bm{\beta}^{(i)},\bm{\theta}^{(i)}\right)}
    \times \frac{\left[\prod_{j=1}^{J-1} f\left(\beta_{kj}'\right)\right]\left[\prod_{j=2}^{J-1} f\left(\theta_j^{(i)} \mid \theta_{j-1}^{(i)},\bm{\beta}_{j}',\bm{\beta}_{j-1}' \right)\right]}{\left[\prod_{j=2}^{J-1} f\left(\theta_j^{(i)} \mid \theta_{j-1}^{(i)},\bm{\beta}_{j}^{(i)},\bm{\beta}_{j-1}^{(i)} \right)\right]}  \nonumber \\
    && \qquad \qquad  \times \frac{1}{q_u\left(u_1,\dots,u_{J-1}\right)} \times \frac{1}{1-p_{\mbox{\scriptsize PO}}} \Biggr].
\end{eqnarray}
We let $u_j$ be i.i.d. random variables such that $u_j \sim N\left(0,\sigma_{P\beta}^2\right)$, where $\sigma_{P\beta}^2$ is the proposal variance. To remove a variable that is currently included we can simply reverse this process, amending the acceptance probabilities accordingly.

%s4.2 ###
\subsection{Updating included variables}
\label{sec:updates}
The second stage of our MCMC routine involves updating the values for any parameters that are currently included in the model. In a random order, we select each of the $K$ variables in turn, and with probability $p_{\mbox{\scriptsize move}}$ we propose new values for the associated parameter(s), and with probability $1-p_{\mbox{\scriptsize move}}$ we propose a shift from PO $\to$ NPO (if variable $k$ has a PO structure), or NPO $\to$ PO (if variable $k$ has an NPO structure).

If variable $k$ has a PO structure, then to update the value of $\beta_k$ we simply propose a new value from some proposal distribution with p.d.f. $q_{\beta }\left(\beta_k^{(i)}\right)$. The update is then a standard Metropolis-Hastings step. Likewise for $\beta_{kj}$ ($j=1,\dots,J-1$) if variable $k$ has an NPO structure.

To switch structures we require a reversible-jump step. To make an NPO $\to$ PO move---i.e. map $\left(\beta_{k1},\dots,\beta_{k(J-1)}\right) \to \beta_k$---we define a bijective function
%e30 ###
\begin{eqnarray}
\label{eq:gNPOtPO}
    g_{NPO \to PO} \left(\beta_{k1}, \dots, \beta_{k(J-1)}\right) &=& \left(\bar{\beta}_{k\cdot},\bar{\beta}_{k\cdot}-2\beta_{k2},\dots,\bar{\beta}_{k\cdot}-2\beta_{k(J-1)}\right) \nonumber \\
    &=& \left(\beta_k,u_1,\dots,u_{J-2}\right),
\end{eqnarray}
where $\bar{\beta}_{k\cdot} = (J-1)^{-1} \sum_{j=1}^{J-1} \beta_{kj}$. To make the reverse move we do not have to propose any new values, and simply use the inverse function
%e31 ###
\begin{eqnarray}
\label{eq:gPOtNPO}
    && g_{PO \to NPO} \left(\beta_k,u_1,\dots,u_{J-2}\right) \nonumber\\
    \qquad &=& \frac{J-1}{2}\left(\beta_k\left(4-J\right)+\sum_{j=1}^{J-2} u_j,\left(\beta_k-u_1\right),\dots,\left(\beta_k-u_{J-2}\right)\right) \nonumber \\
    \qquad &=& \left(\beta_{k1}, \dots, \beta_{k(J-1)}\right).
\end{eqnarray}
These choices are based around the moment matching approach of~\cite{brooksetal:2003}. The acceptance probability for a PO $\to$ NPO move is:
%e32 ###
\begin{eqnarray}
\label{eq:accPOtNPO}
    &&\alpha = \min \Biggl[ 1,\frac{f\left(\bm{Y} \mid \bm{\beta}',\bm{\theta}^{(i)}\right)}{f\left(\bm{Y} \mid \bm{\beta}^{(i)},\bm{\theta}^{(i)}\right)}
     \times \frac{\prod_{j=1}^{J-1} f\left(\beta_{kj}'\right)}{f\left(\beta_k^{(i)}\right)}\nonumber\\
     &&\qquad \qquad  \times \frac{\prod_{j=2}^{J-1} f\left(\theta_j^{(i)} \mid \theta_{j-1}^{(i)},\bm{\beta}_{j}',\bm{\beta}_{j-1}' \right)}{\prod_{j=2}^{J-1} f\left(\theta_j^{(i)} \mid \theta_{j-1}^{(i)},\bm{\beta}_{j}^{(i)},\bm{\beta}_{j-1}^{(i)} \right)} \times \frac{1}{q_u\left(u_1,\dots,u_{J-2}\right)} \nonumber \\
     && \qquad \qquad \times \left(J-1\right)\left(\frac{J-1}{2}\right)^{J-2}
        \Biggr],
\end{eqnarray}
where the final term is the absolute value for the determinant of the Jacobian. Similarly, the acceptance probability for an NPO $\to$ PO move is
%e33 ###
\begin{eqnarray}
\label{eq:accNPOtPO}
    &&\alpha = \min \Biggl[ 1,\frac{f\left(\bm{Y} \mid \bm{\beta}',\bm{\theta}^{(i)}\right)}{f\left(\bm{Y} \mid \bm{\beta}^{(i)},\bm{\theta}^{(i)}\right)}
     \times \frac{f\left(\beta_k'\right)}{\prod_{j=1}^{J-1} f\left(\beta_{kj}^{(i)}\right)}\nonumber\\
     &&\qquad \qquad  \times \frac{\prod_{j=2}^{J-1} f\left(\theta_j^{(i)} \mid \theta_{j-1}^{(i)},\bm{\beta}_{j}',\bm{\beta}_{j-1}' \right)}{\prod_{j=2}^{J-1} f\left(\theta_j^{(i)} \mid \theta_{j-1}^{(i)},\bm{\beta}_{j}^{(i)},\bm{\beta}_{j-1}^{(i)} \right)} \times \frac{q_u\left(u_1,\dots,u_{J-2}\right)}{1} \nonumber\\
     && \qquad \qquad  \times \left(\frac{1}{J-1}\right)\left(\frac{2}{J-1}\right)^{J-2}
        \Biggr].
\end{eqnarray}
We then proceed to update the cut-points, $\bm{\theta}^{(i)}$, in the same way as described in Section~\ref{sec:NPO}.

We note that this general RJ-MCMC algorithm can be adapted in various ways simply by altering the move probabilities. For example, we can remove the variable selection steps and just allow the model to move betwen the PO and NPO structures for each variable by setting $p_{\mbox{\scriptsize jump}}=0$. Similarly, we can also fix all parameters to have either a PO or NPO structure, both with or without variable selection, by adjusting $p_{\mbox{\scriptsize jump}}$ and $p_{\mbox{\scriptsize move}}$ accordingly.

%s4.3 ###
\subsection{Tuning}
In some cases there may be some identifiability issues between regression parameters and their corresponding inclusion indicators when implementing variable selection routines using the framework decribed above. For example, there may be almost identical likelihoods when a parameter is removed (set to zero) and when a parameter is present but has a value close to zero (e.g.~\citealp{ohara_sillanpaa:2009}). If vague priors are used for the regression parameters and inclusion indicators, then these parameters may be unidentifiable. One way to control this is to use a more informative prior, such as one guided by the data or training runs of the model. However, as~\cite{ohara_sillanpaa:2009} note, there is a danger that these approaches will contravene the philosophical construct that the prior distribution should represent one's beliefs about the parameters \textit{before} obtaining any data.

A potential way to tackle this problem in this case is to introduce a hyperprior governing the variance component of the priors for the regression parameters, $\bm{\beta}$. This could be done in various ways, but for PO structures we set
%e34 ###
\begin{equation}
    \beta_k \sim N\left(0,\sigma_{k\beta}^2\right)~\mbox{where}~\sigma_{k\beta} \sim U\left(0,\xi\right)
\end{equation}
and $\xi$ is the maximum \textit{a priori} range for $\sigma_{k\beta}$~\citep{ohara_sillanpaa:2009}, and for NPO structures we set
%e35 ###
\begin{equation}
    \beta_{kj} \sim N\left(0,\sigma_{kj\beta}^2\right)~\mbox{where}~\sigma_{kj\beta} \sim U\left(0,\xi\right).
\end{equation}
This adds a further complexity to the model since it introduces additional parameters to sample during the dimension-jumping steps. For example, a 0 $\to$ PO move would now consist of moving from 0 $\to$ $\left(\beta_k,\sigma_{k\beta}\right)$, likewise a PO $\to$ NPO jump would consist of moving from $\left(\beta_k,\sigma_{k\beta}\right) \to \left(\beta_{k1},\sigma_{k1\beta},\dots, \right.$ $\left.\beta_{k(J-1)},\sigma_{k(J-1)\beta}\right)$ and so on. To do this we update each $\beta$ parameter and its corresponding $\sigma_{\beta}$ parameter at the same time, using independent proposal distributions.

A slight complexity is that the standard deviations must be positive. Hence for a 0 $\to$ PO or 0 $\to$ NPO move (or the reverse moves), we use the same bijective functions as are described in Section~\ref{sec:addrem}, except that the dummy variables for the standard deviations are i.i.d. samples from a $U\left(0,\xi\right)$ distribution. The acceptance probabilities are adjusted accordingly. For a PO $\to$ NPO, we use a slightly different bijective function for proposing the standard deviations than for the regression parameters. Here we propose values for $u_1,\dots,u_{J-2}$ as i.i.d. $U\left(\max\left[0,\sigma_k-\tau_{\sigma}\right],\min\left[\sigma_k+\tau_{\sigma},\xi\right]\right)$ variables, and then define
%e36 ###
\begin{eqnarray}
\label{eq:gPOtNPOsd}
    g_{PO \to NPO} \left(\sigma_k,u_1,\dots,u_{J-2}\right) &=& \left(\sigma_k,\sigma_k+u_1,\dots,\sigma_k+u_{J-2}\right) \nonumber \\
    &=& \left(\sigma_{k1},\dots,\sigma_{k(J-1)}\right).
\end{eqnarray}
To make the reverse move we do not have to propose any new values, and simply use the inverse function
%e37 ###
\begin{eqnarray}
\label{eq:gNPOtPOsd}
    g_{NPO \to PO} \left(\sigma_{k1}, \dots, \sigma_{k(J-1)}\right) &=& \left(\sigma_{k1}, \sigma_{k2}-\sigma_{k1},\dots,\sigma_{k(J-1)}-\sigma_{k1}\right) \nonumber \\
    &=& \left(\sigma_k,u_1,\dots,u_{J-2}\right).
\end{eqnarray}
The acceptance probabilities are updated accordingly, but the additional proposals of the standard deviation terms do not change the Jacobian terms in (\ref{eq:accPOtNPO}) or (\ref{eq:accNPOtPO}).

%s4.4 ###
\subsection{Including categorical explanatory variables with $>2$ levels, and interaction effects}
When comparing nested models including interaction effects, it is usual to specify that interactions can only be included as long as the corresponding main effect terms are also included, and that higher-order interaction terms are included only if all lower-order terms are included~\citep{krzanowski:1998}. These constraints can be incorporated into the routines described in Section~\ref{sec:RJ} by altering the move probabilities. For example, consider the possible moves for a main effect variable, $X_k$, currently included in the model (with a PO structure). If there were no interaction effects, then we propose to exclude the variable with probability $p_{\mbox{\scriptsize jump}}$. If we are modelling interaction effects, then we would instead propose to exclude the variable with probability $p_{\mbox{\scriptsize jump}}'$, where
%e38 ###
\begin{equation}
    p_{\mbox{\scriptsize jump}}' = \left\{ \begin{array}{ll}
                    0 & \mbox{if any interaction effect relating to $X_k$ is present},\\
                    p_{\mbox{\scriptsize jump}} & \mbox{otherwise}.
                    \end{array}
                    \right.
\end{equation}
Likewise, to add interaction effects we need to check that all associated main effects and lower-order interaction effects are present first. This ensures that we only drop or add variables in the correct manner.

Explanatory variables with $>2$ categories require more than one dummy variable to model (see Section~\ref{sec:criteria}). In this case, when proposing to add or remove a variable of this form, we must ensure that all associated dummy variables are added or removed simultaneously. We propose to add a variable of this nature with probability $p_{\mbox{\scriptsize jump}}$, and then for each associated dummy variable $X_{k}$, we independently propose whether these will have PO or NPO structures on addition, with probabilities $p_{\mbox{\scriptsize PO}}$ or $p_{\mbox{\scriptsize NPO}}$ respectively. The acceptance probabilities are amended accordingly, with the Jacobian term being just a product of the corresponding Jacobian terms for each of the dummy variables. The reverse process proceeds in a similar manner.

%s5 ###
\section{Applications}
\label{sec:app}
All the following routines were coded in C and R~\citep{Rstats} and are
\mbox{available} in an R package called \texttt{BayesOrd}, which in
turn uses the \texttt{coda}~\citep{R_coda} and \texttt{multicore}~\citep{R_multicore}
packages to produce output and run multiple chains in parallel.
All results are reported to 2 significant figures (s.f.). The development~\mbox{version} of this package is available at \href{https://github.com/tjmckinley/BayesOrd}{https://github.com/tjmckinley/BayesOrd}. %The plots in Figure~\ref{fig:sims} are produced using the \texttt{ggplot2} package~\citep{wickham:2009} in R.
Following~\cite{link_eaton:2012}, we do not thin our MCMC chains once the burn-in
has been \mbox{discarded}.\looseness=-1

%s5.1 ###
\subsection{Simulation study}
To test the performance of our algorithms, we simulated different data sets assuming
\begin{enumerate}[(a)]
    \item each variable has a PO structure;
    \item each variable has an NPO structure; and
    \item a mixture of PO and NPO variables are used.
\end{enumerate}
For each scenario we simulated $n_{\mbox{\scriptsize sim}} = 100$ data sets, each containing $n = 1000$ samples. Each sample corresponds to measurements on the response variable ($Y$) and 7 explanatory variables (5 binary, $X_1, \dots, X_5$, and 2 discrete, $X_6$ and $X_7$). The response is an ordinal variable with three levels.

Each simulation proceeds as follows:
\begin{enumerate}
    \item In scenario (a), set each explanatory variable, $X_k$, to have a PO structure. In scenario (b) set each structure to NPO, and in scenario (c) sample the structure for each $X_k$ from a Bernoulli distribution with probability 0.5.
    \item For each categorical variable $X_{ik}$ ($i=1, \dots, n$; $k = 1, \dots, 5$), sample its value (0 or 1) from a Bernoulli distribution with probability 0.5. (Ensure that there are at least 5\% of samples in each group by resampling if required.)
    \item For each discrete $X_{ik}$ ($i=1, \dots, n$; $k = 6, 7$), sample data points as $X_{ik} = |X_{ik}'|$, where $X_{ik}' \sim N\left(0, \sigma_k^2\right)$. Here, $\sigma_k = |\sigma_k'|$ and $\sigma_k' \sim N(0, 5^2)$.
    \item Sample the regression parameters $\beta_{kj} \sim N(0, 5^2)$, where $j = 1, 2$ corresponds to the length of the response. For each $k$ corresponding to a PO structure, set $\beta_{kj} = \beta_{k1} \forall j$.
    \item Sample the first threshold parameter, $\theta_1 \sim N(-1, 0.1^2)$, and then simulate the second threshold parameter, $\theta_2$ conditional on $\theta_1$, the simulated data $\bm{X}$ and the regression parameters \bm{\beta}, ensuring that the stochastic ordering conditions (\ref{eq:stochorder}) hold. To do this we can use (\ref{eq:nonPOcond2}) to define a lower bound for $\theta_2$, and then add some positive random noise (we chose the absolute value from a $N\left(0, 0.1^2\right)$ distribution). (Note that in the case of scenario (a) we only need to simulate such that $\theta_1 < \theta_2$, since the stochastic ordering conditions always hold.)
    \item Finally, sample values of the response variable, $Y_i$, from a multinomial distribution with probability vector defined using (\ref{eq:PPOeq}). (Ensure that there are at least 5\% of the samples in each category of the response, else re-simulate.)
\end{enumerate}

Once the data sets were simulated, we proceeded to fit PO and NPO models in both Bayesian and maximum likelihood (ML) frameworks. The Bayesian models were fitted using the routines developed in this manuscript and implemented in the \texttt{BayesOrd} package. The maximum likelihood PO models were fitted using the \texttt{polr} function in the \texttt{MASS}~\citep{venables_ripley:2002} package in R, and the ML NPO models were fitted using binary logistic regressions, as described in e.g.~\cite{bender_grouven:1998}. We also fitted a Bayesian PPO model, using the reversible-jump routines described earlier to choose between the competing structures for each variable. For the MCMC routines, we used 200,000 updates, with the first 10,000 discarded as burn-in.

To summarise the results we examine the distributions for the squared error between the true value of the regression parameters and the ML estimate or posterior mean accordingly. Table~\ref{tab:sims} summarises these results. Focussing first on the results from scenario (a), we can see that as expected, the PO models perform well, with the ML and Bayesian estimates showing a similar degree-of-accuracy. The NPO and PPO models also perform well, suggesting that although they are overparameterised, given enough data they can produce robust inference on the parameters.

%On occasion these models fit the data poorly (shown by the large maximum value and consequently the skewed means). These are not so obvious in the Bayesian results as in the ML results, because the use of priors constrains the degree of lack-of-fit. However, examining the trace and density plots for these fits highlights the lack of identifiability, and as such care must be taken in comparing these extreme values, which would almost certainly be picked up during model diagnostic steps.
%It certainly seems that for the odd simulation the ML approach does not fit very well (shown by the large maximum value and consequently the skewed means), and the Bayesian approaches seem more robust in this sense.

For data sets simulated using scenario (b), the PO models now fit poorly, but the NPO and PPO models once again perform well. Similar patterns are observed for the simulations based on scenario (c), and once again the Bayesian NPO and PPO models perform very well in comparison to the other approaches. There are occasional poor estimates of the parameters used in the simulations (seen by the high 97.5\% credible intervals in Table~\ref{tab:sims}). These could be caused either by a lack-of-fit (most likely when these values are very high), or more frequently when the data are a sample from the extremes of the expected sampling distribution. In any case, the Bayesian methods seem more robust to these outliers, particularly compared to the extreme ML NPO mismatches. We postulate that this is likely due to the fact that the Bayesian methods contain all information in the likelihood, as opposed to the ML NPO method which must treat groups independently.

As a simple exploration of the utility of the Bayesian PPO model for discriminating between PO and NPO structures, we apply a threshold such that any variable with $\mbox{PPA}_{\mbox{\scriptsize PO}} > 0.5$ is classified as having a PO structure. In this case we have $7 \times 100$ possible predictions for each simulation scenario. In the case of scenario (a), only $3/700 = 0.43\%$ are incorrectly specified as having an NPO structure. In the case of scenario (b), $229 / 700 = 33\%$ are misclassified as having a PO structure. For scenario (c), $124/700 = 18\%$ are misclassified, of which $95/700 = 14\%$ are NPO variables misclassified as PO variables, and $29/700 = 4\%$ are PO variables misclassified as NPO variables. This shows a good predictive power, bearing in mind that the Bayesian model choice framework intrinsically favours more parsimonious models, and as such the majority of misclassifications were NPO variables being reduced to PO variables, such as we might expect if the differences between the regression parameters for each level of the response are small. Of course these `misclassifications' may be directly due to quirks in the data as a result of random sampling, and to this end Table~\ref{tab:sims} suggests that the Bayesian NPO and PPO estimates are robust compared to other methods, even accounting for any misclassification in the actual structure used for the simulations. We reiterate that these model fits were performed blind, without a prerequisite descriptive analysis that might shed some light on our \textit{a priori} expectations of variable structures. In practice we would take more care with our preliminary model exploration and our model diagnostics, but with this in mind we think the methods perform well.
%t1 ###
\begin{table}[!ht]
    \centering
    \begin{tabular}{|l|llll|}
        \hline
        \textbf{Scenario} & \textbf{Algorithm} & \textbf{2.5\%} & \textbf{Median} & \textbf{97.5\%} \\
        \hline
        \multirow{5}{*}{(a) PO} & ML (PO) & 0.00 & 0.06 & 3.5 \\
        & Bayesian (PO) & 0.00 & 0.06 & 1.8 \\
        \cline{2-5}
        & ML (NPO) & 0.00 & 0.10 & 7.8 \\
        & Bayesian (NPO) & 0.00 & 0.07 & 3.1 \\
        \cline{2-5}
        & Bayesian (PPO) & 0.00 & 0.06 & 1.9 \\
        \hline
        \multirow{5}{*}{(b) NPO} & ML (PO) & 0.02 & 5.2 & 85 \\
        & Bayesian (PO) & 0.02 & 6.3 & 86 \\
        \cline{2-5}
        & ML (NPO) & 0.00 & 0.32 & 1821 \\
        & Bayesian (NPO) & 0.00 & 0.37 & 34 \\
        \cline{2-5}
        & Bayesian (PPO) & 0.00 & 0.48 & 37 \\
        \hline
        \multirow{5}{*}{(c) PPO} & ML (PO) & 0.00 & 2.5 & 61 \\
        & Bayesian (PO) & 0.00 & 3.4 & 66 \\
        \cline{2-5}
        & ML (NPO) & 0.00 & 0.24 & 321 \\
        & Bayesian (NPO) & 0.00 & 0.18 & 18 \\
        \cline{2-5}
        & Bayesian (PPO) & 0.00 & 0.16 & 20 \\
        \hline
    \end{tabular}
    \caption{Summaries of squared error between estimated and true values, for data sets generated using three different scenarios (defined in the main text). Within each panel, $n_{\mbox{\scriptsize sim}} = 100$ simulated data sets are generated, each of size $n = 1000$ samples. Each panel is further stratified by the type of model (PO, NPO, PPO) and the fitting mechanism (ML or Bayesian).\label{tab:sims}}
\end{table}
%   \begin{center}
%   \makebox{\includegraphics[width = \textwidth]{simcomparisons.eps}}
%   \end{center}
%   \caption{Summaries of squared error between estimated and true values, for $n_{\mbox{\scriptsize sim}} = 100$ simulated data sets. Each plot is stratified by the type of model and fitting mechanism. The plots correspond to a) all variables; b) just variables with a true PO structure; c) just variables with a true NPO structure; d) the subset of variables with a true PO structure that are mis-classified by the Bayesian PPO model, and e) the subset of variables with a true NPO structure that are mis-classified by the Bayesian PPO model. (Note that the y-axes have been truncated at 100 for clarity.)}
%   \label{fig:sims}

%s5.2 ###
\subsection{Longitudinal study of individual-level risk factors affecting body condition score in a dog population in Zenzele, South Africa}
These data form part of a wider study to examine the impact of immunological and demographic factors on canine rabies vaccination coverage.  This study was conducted in four locations: Braamfischerville and Zenzele in Gauteng province, South Africa; and Antiga and Kelusa in Bali province, Indonesia. Full details of the study, and a comprehensive analysis of all the data collected from each of the sites is provided in~\cite{mortersetal:2014}.

\begin{figure}
    \makebox{\includegraphics{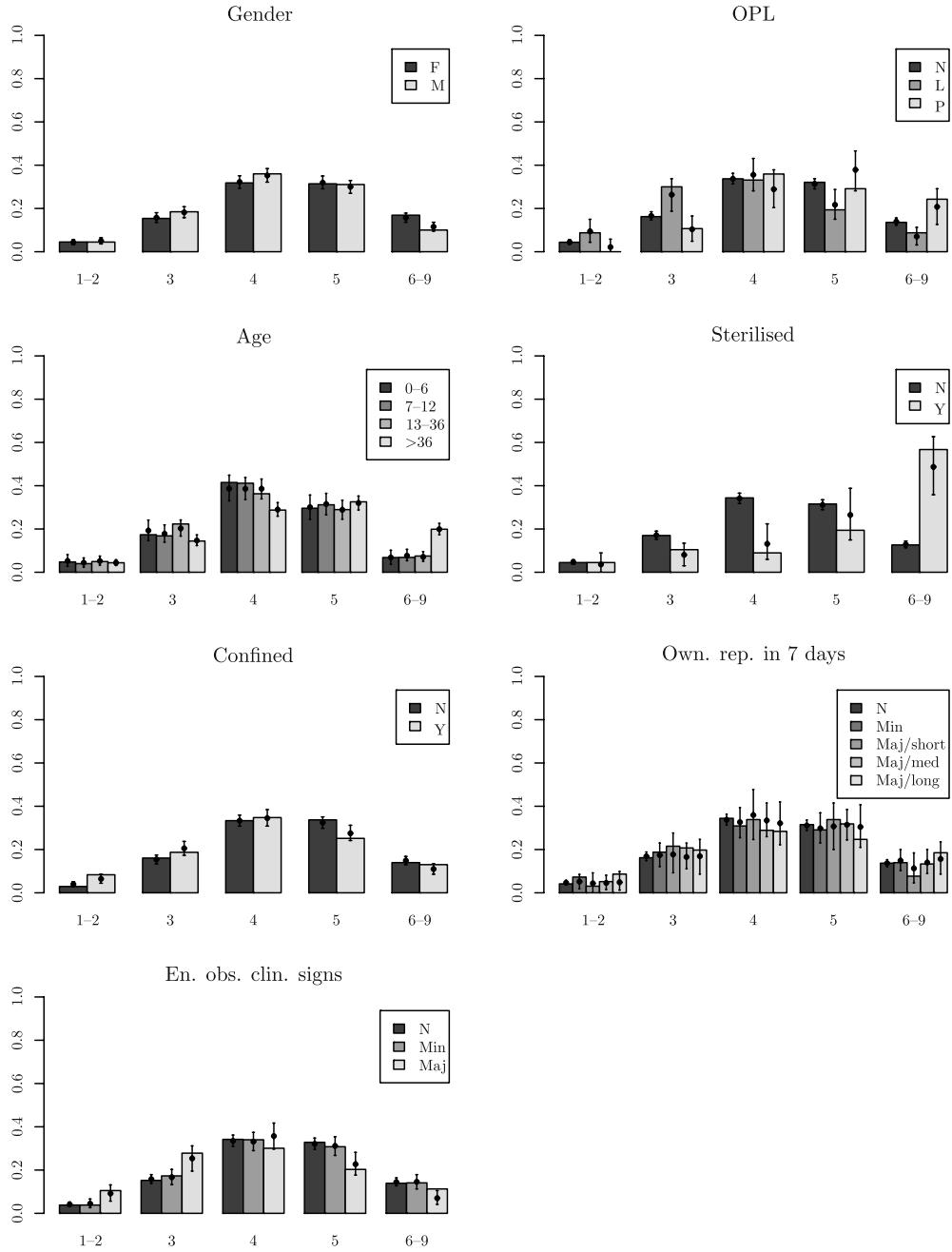}}
    \caption{Marginal posterior predictive distributions for the explanatory variables, against the observed data. Bars represent the data, the points are the marginal predictive means and the error bars are the 95\% prediction intervals.}
    \label{fig:fitted}
\end{figure}

To illustrate the methodology, we focus attention on one particular data set from Zenzele, exploring individual-level risk factors associated with body condition score in a population of dogs. The data set consists of 2746 entries, for 738 dogs, with each dog examined between 1 and 17 times across the period 3rd March 2008--8th April 2011. Body condition score (BCS) was assessed using a nine-point scoring system~\citep{german_holden:2006}, with each dog being scored by two assessors simultaneously. The system assigns a score of 1--9, with 1 being very underweight, 5 being normal, and 9 being obese. To maintain a reasonable sample size in each group, we amalgamated the extreme scores, resulting in 5 BCS groups: 1--2, 3, 4, 5 and 6--9. Eight explanatory variables were collected: gender (male/female), OPL (oestrus-pregnancy-lactation; coded as normal/lactating/pregnant), number of dogs in the sample unit (discrete between 1--9), age (0--6 months, 7--12 months, 13--36 months and $>$36 months), sterilisation (true/false), confinement (true/false), owner reported clinical signs in the previous 7 days (none/minor/major-short duration/major-medium duration/major-long duration) and clinical signs observed by enumerator during interview (none/minor/major). In the interests of comparison, we fitted two separate models, the first assuming that the \textit{maximum} BCS between the two assessors was correct, and the second assuming the \textit{minimum} was correct. Summaries of the data are provided in Table~\ref{tab:data}, and distributions by BCS are shown in Figure~\ref{fig:fitted}.

%t2 ###
\begin{table}[!ht]
    \begin{center}
    \begin{tabular}{|c|c|c|}
    \hline
    \textbf{Variable} & \textbf{Level} & \textbf{Count / summary}\\
    \hline
    \multirow{5}{*}{BCS} & 1--2 & 123\\
    & 3 & 462\\
    & 4 & 927\\
    & 5 & 858\\
    & 6--9 & 376\\
    \hline
    \multirow{2}{*}{Gender} & Female & 1468\\
    & Male & 1278\\
    \hline
    \multirow{3}{*}{OPL} & Normal & 2483\\
    & Lactating & 160\\
    & Pregnant & 103\\
    \hline
    \multirow{6}{*}{\# dogs in SU} & & \textit{Min}: 1\\
     & & \textit{Lower quartile}: 1\\
     & & \textit{Median}: 1\\
     & & \textit{Mean}: 1.8\\
     & & \textit{Upper quartile}: 2\\
     & & \textit{Max}: 9\\
    \hline
    \multirow{4}{*}{Age} & 0--6m & 294\\
    & 7--12m & 452\\
    & 13--36m & 587\\
    & $>$ 36m & 1413\\
    \hline
    \multirow{2}{*}{Sterilisation} & No & 2679\\
    & Yes & 67\\
    \hline
    \multirow{2}{*}{Confinement} & No & 1955\\
    & Yes & 791\\
    \hline
    & None & 2300\\
    Owner reported & Minor & 165\\
    clinical signs in & Major/short & 65\\
    previous 7 days & Major/med. & 135\\
    & Major/long & 81\\
    \hline
    Clinical signs & None & 1983\\
    observed by & Minor & 497\\
    enumerator & Major & 266\\
    \hline
    \end{tabular}
    \caption{Marginal summaries of the data (assuming \textit{maximum} BCS). The final column contains counts unless otherwise stated. For comparison, the BCS counts when choosing the \textit{minimum} BCS are 191, 652, 1082, 617 and 204 respectively.\label{tab:data}}
    \end{center}
\end{table}

To account for the repeated measurements, an individual dog-level term, $\psi_{D_i}$, was introduced, with prior distribution
%e39 ###
\begin{equation}
    \psi_{D_i} \sim N\left(0,\sigma_{\psi}^2\right),
\end{equation}
where $D_i$ denotes the specific dog corresponding to observation $i$ ($D_i=1,\dots,738$), and $\sigma_{\psi}^2$ has a vague gamma hyperprior with shape and rate parameters 0.01 and 0.01 respectively (i.e. mean=1 and variance=100). At each iteration 30\% of the $\psi_{D_i}$ terms were updated in turn at random, using a uniform random walk proposal with the maximum proposal jump given by $\tau_{\psi}$. Likewise the variance $\sigma_{\psi}^2$ was also updated in the same manner with the maximum proposal jump given by $\tau_{\sigma_{\psi}}$.

To complete the Bayesian specification we set the prior variance for the cut-points, $\sigma_{\theta}^2=1$, and the maximum \textit{a priori} value for the standard deviations of the regression parameter priors, $\xi=20$ (following~\citealp{ohara_sillanpaa:2009}). The proposal parameters were: $\tau_{\beta}=1$, $\tau_{\theta}=1$, $\sigma^2_{P\beta}=1$, $\tau_{\sigma}=1$, $\tau_{\psi}=1$ and $\tau_{\sigma_{\psi}}=1$. Two chains were run, and after a short training run of 1,000 iterations, from which initial values for the main chains were generated, we ran 500,000 iterations with the first 50,000 discarded as burn-in. To produce the fitted plots (Figure~\ref{fig:fitted}) we took 2,000 samples from the posterior. Full trace and density plots are given in Supplementary Materials.

The PPAs for each variable, averaged across the competing models are shown in Table~\ref{tab:PPAs}. For clarity, values $< 1 \times 10^{-1}$ are rounded to zero. More precise results are shown in Supplementary Table S1. Using conventional rules-of-thumb for interpreting these values (see e.g.~\citealp{viallefontetal:2001}), if a variable has a PPA of inclusion of $<$0.5, then we consider that there is negligible evidence to support this variable being associated with the response. PPAs of inclusion of 0.5--0.75 are considered weak evidence, 0.75--0.95 positive evidence, 0.95--0.99 strong evidence and $>$0.99 very strong evidence.

%t3 ###
\begin{table}[!ht]
    \begin{center}
    \begin{tabular}{|cc|ccc|ccc|}
        \hline
        &&\multicolumn{3}{|c|}{\textbf{Max. BCS}}&\multicolumn{3}{|c|}{\textbf{Min. BCS}}\\
        \hline
        \textbf{Variable}&\textbf{Level}&\textbf{PO}&\textbf{NPO}&\textbf{Exc.}&\textbf{PO}&\textbf{NPO}&\textbf{Exc.}\\
        \hline
        \multirow{2}{*}{Gender}&F&&&&&&\\
        &M&0.52&0&0.48&0.72&0&0.28\\
        \hline
        \multirow{2}{*}{OPL}&Norm.&&&&&&\\
        &Lac.&1&0&0&1&0&0\\
        &Preg.&1&0&0&1&0&0\\
        \hline
        \# dogs in SU&&0.025&0&0.98&0.038&0&0.96\\
        \hline
        \multirow{4}{*}{Age}&0--6m&&&&&&\\
        &7--12m&1&0&0&1&0&0\\
        &13--36m&1&0&0&1&0&0\\
        &$>$36m&0&1&0&0.82&0.18&0\\
        \hline
        \multirow{2}{*}{Sterilised}&N&&&&&&\\
        &Y&0.5&0&0.5&0.43&0&0.57\\
        \hline
        \multirow{2}{*}{Confined}&N&&&&&&\\
        &Y&0.98&0.02&0&1&0&0\\
        \hline
        &None&&&&&&\\
        Owner reported&Minor&0&0&1&0&0&1\\
        clinical signs in&Major/short&0&0&1&0&0&1\\
        previous 7 days&Major/med.&0&0&1&0&0&1\\
        &Major/long&0&0&1&0&0&1\\
        \hline
        Clinical signs&None&&&&&&\\
        observed by&Minor&1&0&0&1&0&0\\
        enumerator&Major&1&0&0&1&0&0\\
        \hline
    \end{tabular}
    \caption{Posterior probabilities of association for different variables, averaged across all models.\label{tab:PPAs}}
    \end{center}
\end{table}

%f1 ###

In this case we can see that there is consistency in the variables identified as being important from both analyses (i.e. using the maximum and minimum BCS scores as the response). In this case we identify gender as showing weak evidence of an association; and OPL, age, confinement and enumerator observed clinical signs as showing very strong evidence of an association.

For those variables with PPAs $>$ 0.5, we can see that in almost all of these cases the PO structure is preferred, which is reflected in the model averaged log cumulative odds ratios shown in Table~\ref{tab:res}. For clarity, where the posterior means and SDs are the same across the levels (to 2 s.f.---i.e. the variable has an effective PO structure), we show only a single result. For all intents and purposes the only variable that shows any possible non-negligible support for an NPO structure is the $>$36 month age class. Posterior predictive distributions for the observed values can be obtained, and the marginal means and 95\% prediction intervals for the categorical explanatory variables are shown in Figure~\ref{fig:fitted}.

%For those variables with a non-zero probability of exclusion (OPL, sterilisation and confinement), these model averaged values will have been shrunk towards zero as a result of the exclusions.

The overall patterns using both the minimum BCS and maximum BCS as response are the same, and so in the following discussion we will focus on the estimates obtained from using the maximum BCS only. When assessing the posteriors, it is possible to produce conditional inference based on a given model, or produce a posterior based on a weighted mixture of the posteriors from each of the models being averaged over (see e.g.~\citealp{kass_raftery:1995,viallefontetal:2001,ohara_sillanpaa:2009}). In the case of variables that have a non-zero posterior probability of exclusion, the latter approach will shrink these estimates towards zero.

With this in mind, males are on average 1.2 times more likely to have a \textit{lower} BCS than females. However, lactating females are, on average, 3.0 times more likely to have a \textit{lower} BCS than equivalent males and non-pregnant females. Pregnant females on the other hand, are 1.6 times more likely to be have a \textit{higher} BCS. For this variable, there is mixed support for the inclusion of gender with a PO structure, and exclusion altogether. Therefore, in this case these posterior estimates will have been shrunk towards zero relative to the conditional posterior given inclusion. This effect will be minimal for the other variables discussed below which each have a high probability of inclusion.

%t4 ###
\begin{table}[htp]
    \begin{center}
    \begin{tabular}{|c|c|c|cc|cc|}
        \hline
        &&&\multicolumn{2}{|c|}{\textbf{Maximum BCS}}&\multicolumn{2}{|c|}{\textbf{Minimum BCS}}\\
        \hline
        \textbf{Variable} & \textbf{Level of} & \textbf{Level of} & \multirow{2}{*}{\textbf{Mean}} & \multirow{2}{*}{\textbf{SD}} & \multirow{2}{*}{\textbf{Mean}} & \multirow{2}{*}{\textbf{SD}}\\
        \textbf{(baseline level)} & \textbf{variable} & \textbf{response} &&&&\\
        \hline
        \multirow{4}{*}{Gender (F)}&\multirow{4}{*}{M}&1&\multirow{4}{*}{-0.18}&\multirow{4}{*}{0.20}&\multirow{4}{*}{-0.28}&\multirow{4}{*}{0.21}\\
        && 2 &&&&\\
        && 3 &&&&\\
        && 4 &&&&\\
        \hline
        \multirow{8}{*}{OPL (normal)}&\multirow{4}{*}{Lac.}&1&\multirow{4}{*}{-1.1}&\multirow{4}{*}{0.18}&\multirow{4}{*}{-1.3}&\multirow{4}{*}{0.18}\\
        && 2 &&&&\\
        && 3 &&&&\\
        && 4 &&&&\\
        \cline{2-7}
        &\multirow{4}{*}{Preg.}&1&0.46&0.25&\multirow{4}{*}{0.47}&\multirow{4}{*}{0.22}\\
        &&2&0.46&0.22 &&\\
        &&3&0.46&0.22 &&\\
        &&4&0.46&0.22 &&\\
        \hline
        \multirow{12}{*}{Age (0--6m)}&\multirow{4}{*}{7--12m}&1&\multirow{4}{*}{0.12}&\multirow{4}{*}{0.15}&\multirow{4}{*}{0.35}&\multirow{4}{*}{0.16}\\
        && 2 &&&&\\
        && 3 &&&&\\
        && 4 &&&&\\
        \cline{2-7}
        &\multirow{4}{*}{13--36m}&1&\multirow{4}{*}{-0.19}&\multirow{4}{*}{0.15}&\multirow{4}{*}{0.09}&\multirow{4}{*}{0.15}\\
        && 2 &&&&\\
        && 3 &&&&\\
        && 4 &&&&\\
        \cline{2-7}
        &\multirow{4}{*}{$>$36m}&1&-0.0065&0.22&0.64&0.28\\
        &&2&0.27&0.18&0.69&0.2\\
        &&3&0.74&0.17&0.79&0.2\\
        &&4&1.5&0.21&0.86&0.31\\
        \hline
        \multirow{4}{*}{Confined (N)}&\multirow{4}{*}{Y}&1&-0.56&0.14&\multirow{4}{*}{-0.58}&\multirow{4}{*}{0.12}\\
        &&2&-0.55&0.12 &&\\
        &&3&-0.55&0.12 &&\\
        &&4&-0.55&0.13 &&\\
        \hline
        &\multirow{4}{*}{Minor}&1&\multirow{4}{*}{-0.14}&\multirow{4}{*}{0.11}&\multirow{4}{*}{-0.14}&\multirow{4}{*}{0.11}\\
        &&2 &&&&\\
        Clinical signs&&3 &&&&\\
        observed by&&4&&&&\\
        \cline{2-7}
        enumerator&\multirow{4}{*}{Major}&1&\multirow{4}{*}{-1}&\multirow{4}{*}{0.15}&\multirow{4}{*}{-1.3}&\multirow{4}{*}{0.15}\\
        (None)&&2&& &&\\
        &&3&& &&\\
        &&4&& &&\\
        \hline
    \end{tabular}
    \caption{Model averaged posterior means and standard deviations for the log cumulative odds ratios. Only those variables with non-negligible assocation to response (i.e. a PPA of inclusion of $>$0.5) are shown. For clarity, those variables that have the same means and SDs for each level of the response (to 2 s.f.) are shown as a single entry.\label{tab:res}}
    \end{center}
\end{table}

The effect of age is interesting; relative to the 0--6 month category, dogs aged between 7--12 months are 1.1 times more likely to have a \textit{higher} BCS; dogs aged between 13--36 months are 1.2 times more likely to have a \textit{lower} BCS, and as adults ($>$36 months) they are between 1--4.5 times more likely to have a \textit{higher} BCS, depending on the category level (since the adult age class has very strong evidence of an NPO structure). A likely explanation is that this pattern reflects normal morphological variation---generally, as dogs become older their activity levels will decrease, resulting in a general increase in BCS.

An interesting finding in this analysis is that other than the $>$ 36 month age category, all other variables had very strong support for a PO structure (conditional on inclusion). One of the key motivations for the study that generated these data was to examine the hypothesis that these canine populations are regulated by environmental resource constraints (as they would be in wild populations). If this hypothesis is true, then consistent with empirical evidence in other species and ecological theory, the thin dogs should generally be the ones with highest energy requirements (particularly lactating and growing dogs). The marginal distributions shown in Figure~\ref{fig:fitted} provide qualitative evidence against this hypothesis, since whilst on average there was a tendency for lactating dogs to be thinner than non-lactating dogs, overall the body condition distribution for lactating dogs shows that most dogs are in reasonable body condition, with fewer dogs in the extremes. Crucially there are underweight lactating dogs and underweight non-lactating dogs, and there are overweight lactating dogs and overweight non-lactating dogs---consistent with variable food availability most likely from an owner, rather than from the environment (e.g. scavenging). The same is true for young dogs.

A similar argument could be made by examining the evidence for PO versus NPO structures for these key variables (particularly OPL and age). Under the hypothesis of environmental constraints limiting population size, then we might expect lactating and young dogs to be more likely to exhibit an NPO structure, with decreasing negative log-odds ratios with increasing BCS. We do not observe this here. There is strong evidence of an NPO structure for the $>$ 36 month age class, though this is again consistent with the population being `managed', rather than acting as a wild population.%, since in the wild older dogs are likely to have lower BCS due to morphological variations resulting in a decreased ability to scavenge or hunt.

Similar results are obtained for all four study regions. This information has important implications for designing optimal vaccination strategies against rabies in these populations. For full details of the study, and a comprehensive discussion about all the collected evidence, see~\cite{mortersetal:2014}.

Confinement is associated with a \textit{lower} BCS, with confined dogs being 1.7 times more likely to have a lower BCS than unconfined dogs. Although confinement, as defined in this study, was highly variable (with regards to the length of time dogs were confined and the frequency that they were released), in general it was observed that dogs that were tied up were often neglected. See~\cite{mortersetal:2014} for a full discussion on these issues.

Finally, the clinical signs variables cover a wide range of possible conditions. These were classified into `minor' (considered unlikely to cause weight loss, such as localised skin lesions and lameness) and `major' (considered likely to cause weight loss, such as vomiting and lethargy). This variable serves as an indicator of the general health of the dog, and it can be seen that as expected, dogs that show evidence of an ongoing medical condition (that is likely to cause weight loss), are more likely to have \textit{lower} BCS values than their healthy counterparts: 1.2 times more likely for minor ailments and 2.7 times more likely for major ailments. %In~\cite{mortersetal:2014}, these results are discussed in comparison to the results found from the other studies.

%s6 ###
\section{Discussion}
\label{sec:dis}
We have introduced a method for fitting cumulative link ordinal regression models that does not require \textit{a priori} assumptions regarding PO or NPO structures to model the relationship between the response and explanatory variables. For categorical explanatory variables we show how stochastic ordering can be ensured in the case of NPO models, and provide a pragmatic approach to ensuring that stochastic ordering holds for continuous or discrete covariates within the range of the observed data. In addition these approaches can be extended to incorporate variable selection within a Bayesian framework, allowing posterior probabilities of association to be produced for competing models. It is straightforward to include individual-level terms to account for repeated measures, and Bayesian model averaging can to be used to provide weighted PPA estimates for the parameters that account for model uncertainty. We have illustrated the methods on a large-scale real-life data set.

The method uses reversible-jump MCMC to jump between models of differing dimensionality. However, implementational difficulties can exist with this method, particularly when jumping between models where the dimensionality is quite different. We found that the simple proposal mechanisms used throughout the paper worked well for this application and others we have tried. Nonetheless it is likely that specific situations may require more additional tuning (as with any MCMC method). For example, if some of the intervals specified by the stochastic ordering conditions (\ref{eq:stochorder}) are small, and the proposal size, $\tau_{\beta}$ is too large, then for NPO structures this may result in a large proportion of proposed values for the $\beta_{kj}$ parameters being rejected as a result of breaking the prior conditions on stochastic ordering. An alternative would be to sample from some form of truncated distribution, though due to the nature of the constraints, this is not trivial.

Another interesting alternative would be to use some form of \textit{shrinkage} model, where the model is defined as
%e40 ###
\begin{equation}
    \log\left(\frac{\gamma_{ij}}{1-\gamma_{ij}}\right) = \theta_j- \bm{\eta}_j^T\bm{X}_i,
\end{equation}
where $i=1,\dots,n$ and $j=1,\dots,J-1$. The conditional prior distributions for the $\eta_{kj}$ parameters are centred around the corresponding $\beta_k$ with a small prior variance. The $\beta_k$ parameters can be given the same prior distribution as before. In this variation the model does not change dimensionality, and so no reversible-jump step is required. The $\eta_{kj}$ parameters then correspond to the degree to which the parameter estimates deviate away from the proportional odds structure. This idea could also be expanded to incorporate variable selection in various ways (see e.g.~\citealp{ohara_sillanpaa:2009}).

Using single-component updates with simple random-walk proposals can also produce Markov chains that are highly autocorrelated, and thus require a large number of iterations and a lot of thinning. Adaptive proposal mechanisms~\citep{haarioetal:2001,roberts_rosenthal:2009} exist for standard (i.e. non-transdimensional) MCMC, that can automatically tune the proposal distributions to produce much more efficient chains in terms of both convergence and mixing. However, it is not currently understood whether these sorts of approaches hold for transdimensional routines, and this is a key area of ongoing research for those who are developing these methods~\citep{hastie_green:2012}. For the kinds of examples shown in this paper the runtimes required to produce a reasonable number of pseudo-independent samples are not prohibitive, and so we do not worry about this aspect here. It is not the purpose of this paper to provide a catch-all routine that works well in every situation, but rather to provide a flexible method that can be adapted to deal with different situations as required.

%As with any multinomial regression technique, care must be taken when fitting models that have low counts in some of the groups; particularly when using categorical explanatory variables.
We occasionally noticed some identifiability issues when fitting NPO models, predominantly between categorical explanatory variables with low counts in some of the groups, and the cut-off parameters. This can be tackled in two main ways: firstly, the variables can be recategorised to ensure that there is a minimum number of individuals in each group. Secondly, we can start the MCMC routines using more informative initial values. Appealing to the Occam's Razor principal, in this paper we decided to generate initial values by producing a short training run, using a PO model that includes all of the explanatory variables (but ignoring the repeated measures). We then ran the full model using the parameter values from the final iteration of the training run as initial values. A similar approach would be to generate maximum likelihood estimates for the simple PO model and use these as initial values instead.

An example of the utility of these routines is that stochastic ordering can be ensured for continuous/discrete covariates within the range of the observed data. It is theoretically possible to ensure these conditions hold for any finite range of values, if an upper or lower bound was known from sources of information other than the observed data. In any case, if it is of interest to extrapolate beyond the range of the data, then it is possible to use the posterior samples to explore the range of covariate values over which the stochastic conditions will hold---essentially building a posterior distribution for the range of valid values. This could also be used as a form of sensitivity analysis to the model assumptions based on the model fit.

It is also worth noting that although we have illustrated these methods using a logistic link function, the methods are applicable to any monotonically increasing link function (though of course the interpretation of the regression parameters will no longer be in terms of the cumulative odds).

%% Supplement Material %%
\begin{supplement}
\stitle{Supplementary Materials: Bayesian model choice in cumulative link ordinal regression models: an application in a longitudinal study of risk factors affecting body condition score in a dog population in Zenzele, South Africa}
\sdatatype{.zip}
\sfilename{Supplementary Materials-May-05-2014.zip}
\slink[doi]{10.1214/14-BA884SUPP}
\end{supplement}

\newcommand{\noopsort}[1]{}

\begin{acknowledgement}
TJM is supported by Biotechnology and Biological Sciences Research Council grant number BB/I012192/1. MM is supported by a grant from the International Fund for Animal Welfare (IFAW) and the World Society for the Protection of Animals (WSPA), with additional support from the Charles Slater Fund and the Jowett Fund. JW is supported by the Alborada Trust and the RAPIDD program of the Science and Technology Directorate, Department of Homeland Security and the Fogarty International Centre. Thanks to Andrew Conlan and Richard Dybowski for useful discussions, and to the anonymous referees whose comments and suggestions helped greatly improve this manuscript.
\end{acknowledgement}

\end{document}